\documentclass[universe,article,accept,pdftex,oneauthor]{Definitions/mdpi} 

\firstpage{1} 
\pubvolume{1}
\issuenum{1}
\articlenumber{0}
\pubyear{2024}
\copyrightyear{2024}
\datereceived{ } 
\daterevised{ } 
\dateaccepted{ } 
\datepublished{ } 
\hreflink{https://doi.org/} 

\Title{Observability of Acausal and Uncorrelated Optical-Quasar Pairs for Quantum-Mechanical Experiments}

\TitleCitation{Observability of Acausal and Uncorrelated Optical-Quasar Pairs for a Quantum-Mechanical Experiment}

\Author{Eric Steinbring}

\AuthorNames{Eric Steinbring}

\AuthorCitation{Steinbring, E.}

\address{National Research Council Canada, Herzberg Astronomy and Astrophysics, Victoria, BC V9E 2E7, Canada; Eric.Steinbring@nrc-cnrc.gc.ca}

\abstract{Viewing high-redshift sources at near-opposite directions on the sky can assure, by light-travel-time arguments, acausality between their emitted photons. One utility would be true random-number generation, by sensing these via two independent telescopes that each flip a switch based on those latest-arrived colours; for example, to autonomously control a quantum-mechanical (QM) experiment. Although demonstrated with distant quasars, those were not fully acausal pairs, which are restricted in simultaneous view from the ground at any single observatory. In optical light such faint sources also require large telescope aperture to avoid sampling assumptions when imaged at fast camera framerates: either unsensed intrinsic correlations between them or equivalently-correlated noise may ruin the expectation of pure randomness. One such case which could spoil a QM test is considered. Based on that, allowed geometries and instrumental limits are modelled for any two ground-based sites, and their data simulated. To compare, an analysis of photometry from the Gemini twin 8-m telescopes is presented, using archival data of well-separated bright stars, obtained with the instruments `Alopeke (on Gemini-North in Hawai'i) and Zorro (on Gemini-South in Chile) simultaneously in two bands (centred at $562~{\rm nm}$ and $832~{\rm nm}$) with 17 Hz framerate. No flux correlation is found, calibrating an analytic model, predicting where a search at signal-to-noise over 50 at 50 Hz with the same instrumentation can be made. Finally, the software PDQ (Predict Different Quasars) is presented which searches a large catalogue of known quasars, reporting those with brightness and visibility suitable to verify acausal, uncorrelated photons at those limits.}


\begin{document}


\section{Introduction}\label{introduction}

That quantum mechanics (QM) must be incomplete, allowing ``spooky" outcomes requiring either super-luminal signals or hidden variables - essentially, unsensed influences on the outcome - was posited by Einstein, Podolsky and Rosen \cite{Einstein1935}. Bell showed, however, that correlations in a QM experiment could allow tests against such hidden-variable theories \cite{Bell1964}. Repeated Bell-theorem tests have since routinely found QM is correct through tightly and simultaneously restricting the necessary conditions on measurements (e.g. \cite{Rosenfeld2017} and references therein), including the influence of experimenter interaction: the so-called "freedom-of-choice" loophole. One route to this has been to set experimental parameters via photons from astronomical sources \citep{Friedman2013, Gallicchio2014}, requiring that interference in such settings had somehow been orchestrated between distant sources and the Earth-based observer. Proof-of-concept tests using stars within the Milky Way were achieved \cite{Handsteiner2017} forcing any collusion in the outcome back hundreds of years. And extension to quasars \citep{Wu2017, Leung2018, Rauch2018} pushed this to $\sim~{\rm Gyrs}$, combining high redshift $z$ with large angular separation on the sky to increasingly place these outside each others' light cone; where $180^\circ$ apart that would be complete if both sources have $z\geq3.65$ \citep{Friedman2013}. Smaller separations could be compensated by higher source redshifts, although possibly at the cost of the objects being fainter. The motivation to reach this limit is independence of settings triggered via those photons, which are unspoiled by communication. But that is true only if no correlated errors are introduced via scattered photons from the sky or otherwise via optical inefficiencies and detector electronics; and so corrupt their unsychronized fluxes just prior to detection.

To outline the timescales and error-limits necessary to meet, in order that the flux of two quasars can be known truly acausal and also measured as sufficiently uncorrelated, consider the one quasar-based Bell test undertaken so far by Rauch et al. \cite{Rauch2018}. It followed the methodology of Clauser \cite{Clauser1969} where an entangled pair of photons emitted from a central source are split between two optical arms and their polarizations detected at receivers. While those entangled photons were in flight, a switching mechanism at each receiver (also co-located with a telescope) selected between two polarization measurements at pre-fixed relative angles, chosen to test the maximum potentially observable difference from QM. That switch was set via the colour of the most recently detected quasar photon: a dichroic splitting flux into two broad colour-bands that did not systematically favour either selection. Bright pairs were viewed separately via two 4-m class telescopes from Observatorio del Roque de los Muchachos on La Palma. One quasar pair was separated by $73^{\circ}$ on the sky, having $z=0.27$ and $3.91$; another pair $84^{\circ}$ apart with $z=0.96$ and $3.91$.  Both quasar fluxes (in ${\rm photons}~{\rm s}^{-1}$) were brighter than sky-background (${\rm photons}~{\rm s}^{-1}~{\rm arcsec}^{-2}$) and a relative polarization measurement was retained only if both quasar photons arrived within the microseconds while the entangled photons were in flight. Fair sampling was assumed, that is, thousands of individual trials were obtained non-continuously (achieving a duty-cycle up to several seconds apart; a cadence $\sim 1~{\rm Hz}$) during runs lasting 12 and 17 minutes, and so each quasar-photon sample initiating a switch was considered independent, providing no information which could be exploited to predict the next outcome. This tolerates at least one in four correlated switching photons being needed to spoil a Bell test, although possibly as little as 14\% is enough \cite{Friedman2019}. Analysis showed that neither the colours of the two quasars nor the background noise against their detection (i.e. sky-line fluctuation) were correlated among all trials beyond such measurement-error margins.

Optical or near-infrared (NIR) photometry from any single observatory site hinders such a QM-test experimental cadence fast enough to remove the fair-sampling assumption for quasars separated by more than $ 90^\circ$ where {\it both} have $z\geq 3.65$. That is due to the two switching telescopes being at most kilometers apart, and $1~{\rm km}$ is $3~\mu{\rm s}$ in light travel time, i.e. $\sim 1~{\rm MHz}$ rates. However, a typical $z=4$ quasar has a V-band AB-magnitude of about 20 \cite{Flesch2023} (roughly the same as dark sky in ${\rm mag}~{\rm arcsec}^{-2}$) and similar out to 1 micron, which even for an idealized unobstructed 8-m class telescope with perfect throughput (and splitting light into two bandpasses) gives a fluence less than 18 photons in 10 milliseconds under photometric skies but low elevation, at a reasonable observing limit $\leq 1~{\rm mag}$ of extinction. Despite good seeing of $0.8~{\rm arcsec}$ this would be $17.5~{\rm photons}~{\rm s}^{-1}~{\rm arcsec}^{-2}$ against a background just below 2 photons, so an individual exposure signal-to-noise ratio (SNR) over 10 for each quasar. A sampling cadence of $100~{\rm Hz}$ would then already be the fastest possible retaining sufficient SNR against the sky background flux for simultaneous two-band photometry of both quasars while reaching a colour-discrimination per sample of $\sqrt{2}\times 10\%\approx 14\%$ uncertainty; and less well for realistic efficiencies. Finding brighter sources or choosing broader passbands could each double the SNR, but still at 4 orders of magnitude slower than the telescope-to-telescope signalling rate. That is a problem because millisecond and longer-lasting oscillatory correlations might naturally arise due to the self-similar Kolmogorov scaling of wavefront aberrations, particularly at smaller telescope angular separations, without any ability to discriminate those against fluctuations due to the sources themselves. Adaptive optics (AO) may help, as such systems routinely sense these phase errors at kilohertz rates in wavelengths near $500~{\rm nm}$ and utilize their $\lambda^{5/3}$ index-of-refraction behaviour to manipulate them longward of $800~{\rm nm}$ or so, into the NIR: those photons can be redirected with a fast-moving optic. This results in a sharpened image, possibly leaving it diffraction-limited at $1~{\rm kHz}$; more readily improving photometry by factors of 2 or so, and leaving some uncorrected fraction at lower frequencies; (e.g. \cite{Steinbring2014}). But, even with this assistance, a practical sampling cadence is too slow to exclude all higher-frequency intrinsic correlations (or correlated noise) that can provide a prediction for the next experimental outcome: whether that quasar switching-photon sample of either band at each telescope is to fall inside the seeing disk, or not.

Instead, telescopes situated at two well-separated sites could rule out such correlations, because then both high-angular separations up to $180^\circ$ (and so known-acausal sources) and fluctuations in seeing (and therefore source versus background flux per sample) need only be monitored on timescales closer to the Earth light-crossing time, roughly less than $0.04~{\rm s}$ or at $25~{\rm Hz}$, to measure and exclude any influence (or temporally-correlated errors) in their simultaneous photometry. Interestingly, no such case is so far reported, and no monitoring campaign simultaneously viewing quasars outside each observatory's horizon seems to have been undertaken, at any wavelength. For $\gamma$-rays and X-rays, and through to far-ultraviolet this would require a specially-designed spaceborne mission, not yet planned. And the difficulty extends to the radio, as from the ground, dish elevations must still stay above the horizon, regardless of Sun position. From Earth's nightside, optical/NIR telescopes are restricted to separations below two airmasses; incuring about twice the zenith extinction and instrumental zeropoint error for each. Even so, non-AO corrected photometry to within 10\% accuracy is possible for those sources. To first order then, a goal to achieve simultaneous photometry at 10 Hz to 100 Hz-framerates of acausal quasar pairs with two independent telescopes may be within reach. What should they see?

The first step to answering that is to define a plausible correlated signal to look for, sufficient to spoil a Bell test, in order to characterize what might be detected within the observational noise - at best just the instrumental zeropoints - without assuming instrinsic randomness. Second is the characterization of instrumental limits, and demonstrating simultaneous high-framerate optical photometry from two widely-separated sites, setting a benchmark. An active galactic nucleus (AGN) need not be the source. So at least one useful dataset to probe is already available at Gemini: on Maunakea in Hawaii ($19.82^{\circ}$N, $155.47^{\circ}$W, 4213 m) and on Cerro Pachon in Chile ($30.24^{\circ}$S, $70.74^{\circ}$W, 2722 m), that when each viewing a target near zenith, places those $95.5^{\circ}$ apart on the sky. These twin 8.1-m telescopes have operated near-identical high-framerate (up to $67~{\rm Hz}$) optical imagers for several years: `Alopeke and Zorro; and for which a public archive contains some serendipitous, simultaneous observations of bright stars. Although such data do not constitute a QM experiment, they do provide a baseline in devising a future one: at over 10600 km apart, no collusion is possible on timescales less than this distance divided by the speed of light or $l/c\approx0.04~{\rm s}$, which in the restframe at $z=4$ corresponds to $0.2~{\rm s}$.

The next section describes how the geometry of an experiment allowing simultaneous photometry of quasar pairs restricts their best relative SNR; if not due simply to correlated errors, AGN-physics can motivate how, on the short timescales relevant here such an intrinsic signal might arise. Simulated observational data are generated. And a method of sensing that correlation of flux differences will be described, for conditions where point-source photometry has sufficient sampling and sensitivity to reach the zeropoints of two identical instruments, within about 2\% error. This is then extended to any two observing sites on Earth, including paired antipodal ones and to fainter pairs, appropriate for $z=4$ quasars. Following that, the available Gemini dataset is described: 4 observational pairs of bright stars with separations up to nearly $130^{\circ}$ on the sky and at peak SNR consistent with the model; providing a calibration at $17~{\rm Hz}$ sampling rate towards tests with quasars. Software is described which finds those for follow-up tests at over $50~{\rm Hz}$ framerate. Discussion concludes with a short list of potential quasar-pairs, and their upcoming best-observed nights for simultaneous Gemini photometry with `Alopeke/Zorro, reaching the necessary photometric accuracy to prove suitability in a Bell test.

\section{Methods of Finding Uncorrelated, Acausal Quasars}\label{methods}

The approach taken here is to first define what correlation between two sources, whether intrinsic or from correlated noise, could be sufficient to spoil a QM test that employed those. Although it seems an improbable scenario, positing that some acausal-source pair might be found perfectly correlated establishes a falsifiable signal that can be looked for. This allows a well-defined estimate of SNR for the measurement of differences in flux of all photons (both causal and acausal), and producing simulated data. With that, a model of observational noise of the two sites can be used to find, via a catalog of known quasars, useful targets for which to initiate a search, despite expecting a null result.

\subsection{A Plausible Correlated Quasar-Pair Signal to Look For}\label{plausible_signal}

Maybe nature does provide a pair of quasars (call them source A and B) that could not have communicated but always have perfectly correlated fluxes within two colour-filter passbands (labelled ``blue" and ``red"). If they exist, such a correlation can be found, even if the long-term average relative A-B colour is constant. For example, imagine that the flux of A at a particular emission wavelength (within either the blue or red passband) is always anti-correlated with B in a sinusoidal function at constant frequency low enough to be detectable; say, seconds to minutes long. That is at least a plausible scenario, because the rotational periods of AGN accretion disks or related structures in their jets are the source of the emission-line and continuum photons under study here, and reverberation mapping has well established the characteristic sizes of those on the order of light days across; they are known to be variable on those timescales and longer \cite{Li2018, Mudd2018}. But if those two disks have identical rotational periods, common physical scales of smaller substructures should also be correlated on shorter timescales too, certainly less than hours. This would be analogous to soundwaves behaving as a ``siren" with contant cycle or pitch, but fluctuating intensity. All that is necessary for detection and use (and as utilized in the Rauch et al. experiment) is reliably reaching a minimum threshold flux in the blue and red passbands; providing correlated photons to predictably trigger the switches. Ideally, there is always a reliable frequency of signal (oscillating between blue and red) of fixed amplitude.

\begin{figure}
\includegraphics[width=9.5cm]{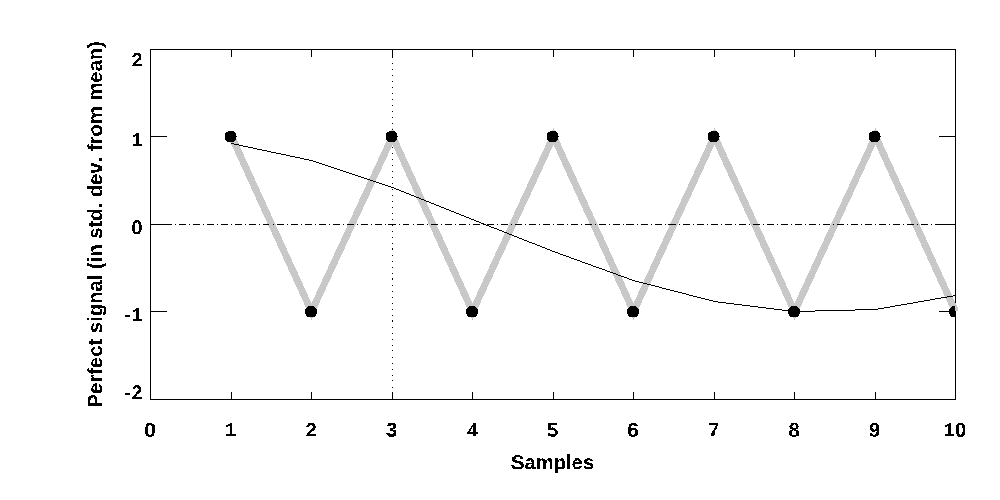}
\caption{A clear signal between two sources A and B: a steady sinusoidal ``siren" with a half-cycle of 8 samples spanning 2 full standard-deviations of each receivers' noise. Minimized to the fastest-attainable framerate, this is a perfect ``toggle-switch" (filled circles, grey line-segments).}\label{figure_schematic_toggle_switch}
\end{figure}

When that frequency increases to the highest framerate possible at the switching telescopes, this amounts to the signals acting as an on/off "toggle switch": equivalently either blue for the observer of A (designated hereafter as Alice) when it is red for observer of B (Bob), or vice-versa - at a constant frequency. This limit is illustrated schematically in Figure~\ref{figure_schematic_toggle_switch}, where the filled-black circles indicate the colour of a received sample of photons being either at least one standard deviation from the mean (long-term average) colour of the received photons at both switching-station detectors. It will become clear later why this particular limit is chosen. But in this scenario, it is obvious that neither Alice nor Bob need know what colour of photons the other received, because when they observe one state (sample 1 is blue) they are confident to two standard deviations of their combined noise sources that the opposite is currently the case for the other (red), and furthermore, that their sample 3 will return to the same state (blue; the other is red again). The longer-term, lower-frequency limit for the detectable siren-like signal (shown for a period of 16 samples) is indicated by a smooth curve; this lower frequency is comparable (for 100 Hz sampling) to the outer-limit of clock precision of either receiving telescope. 

\subsection{Simulated Data With Correlated Signals and Gaussian Noise}\label{autocorrelation}

\begin{figure}
\includegraphics[width=14cm]{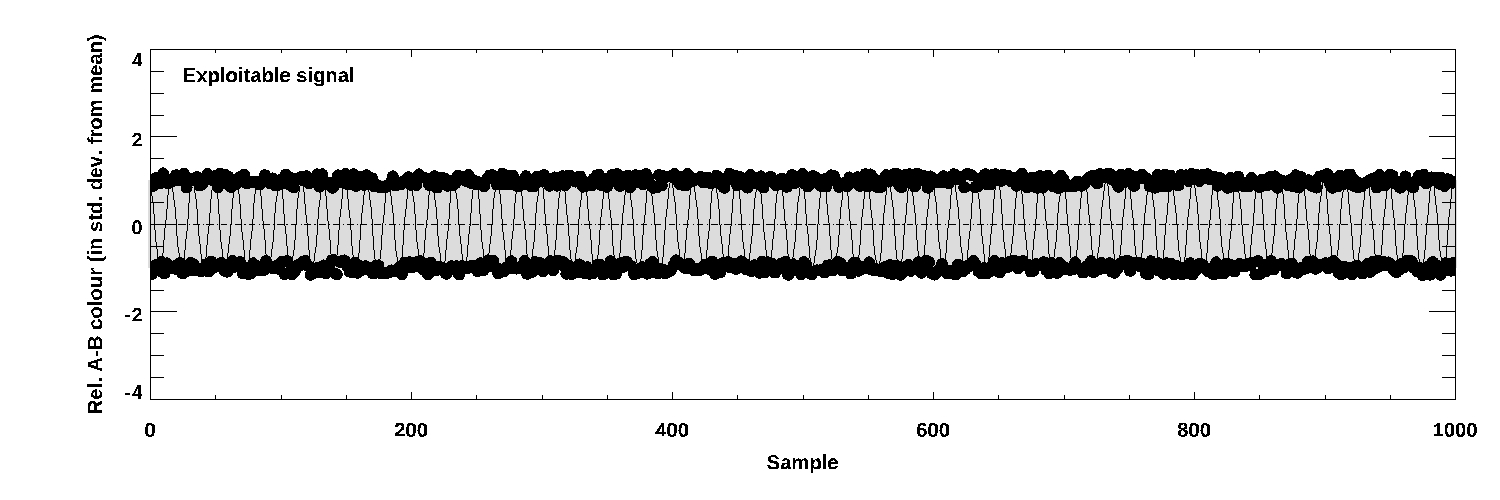}
\includegraphics[width=14cm]{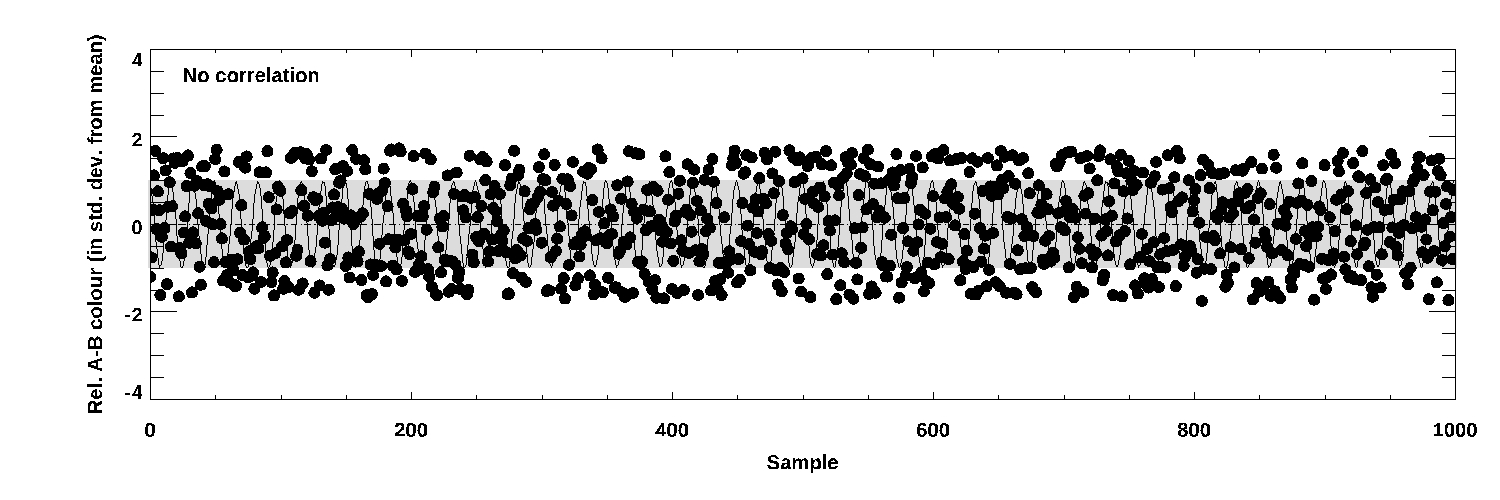}
\includegraphics[width=14cm]{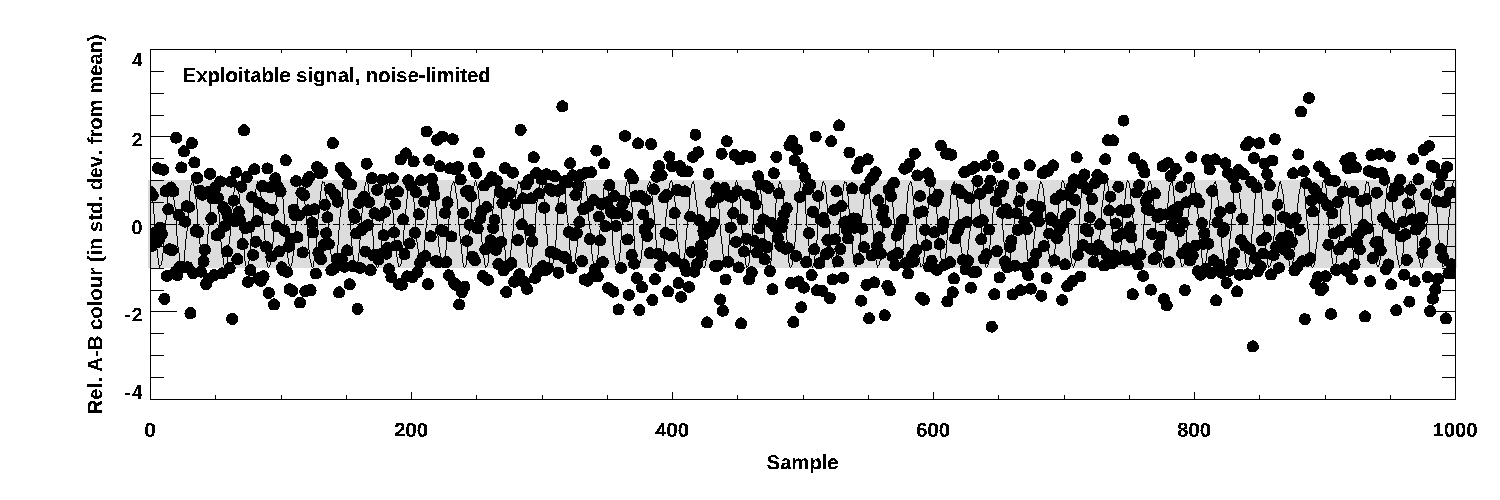}
\includegraphics[width=14cm]{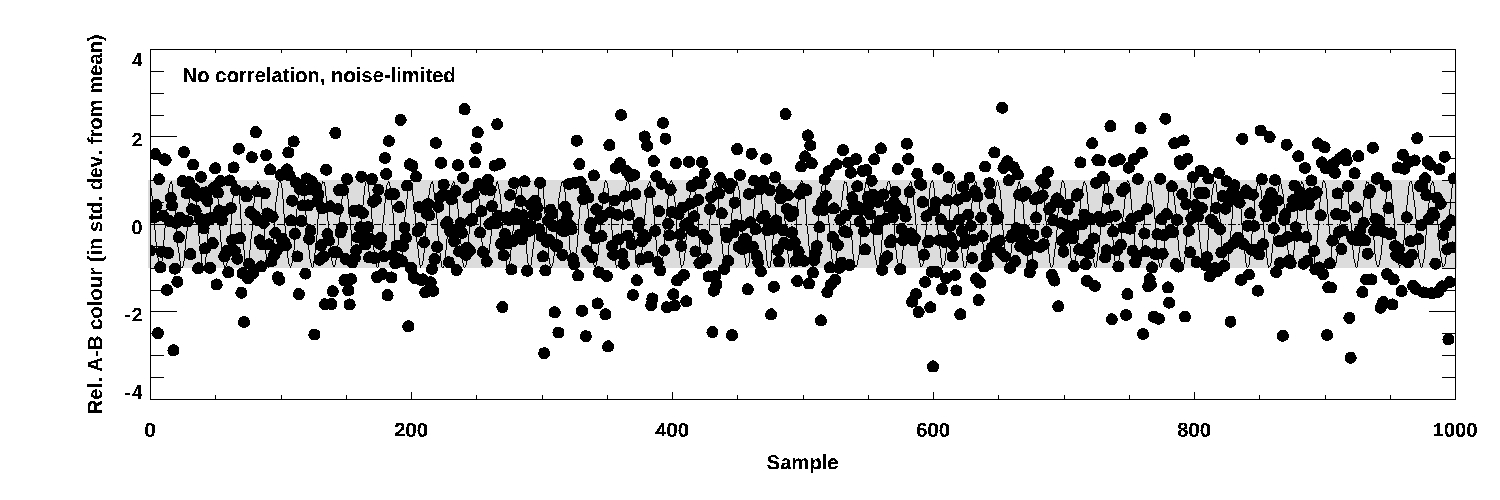}
\caption{Potential minimally ``exploitable" A-B colour signal persisting for 1000 samples (top panel) and spanning necessary one standard-deviation from the median (shaded grey); a thin black curve indicates a perfect sinusoid; or, an equivalent distribution of pure white noise (second-to-top panel). Below are simulated observational sequences: just-sufficient Gaussian noise to void its utility for exploitation in a QM test, or (bottom panel) no signal at all with that same amount of noise.}\label{figure_simulated_relative_colours}
\end{figure}

Simulated data were generated for the A-B colour signal of Subsection~\ref{plausible_signal}, clearly exploitable to predict future settings, but degraded in a way to be only just-sufficient at the receiver/switching stations to spoil a QM test. Figure~\ref{figure_simulated_relative_colours} shows what would be observed if the signal as described in Figure~\ref{figure_schematic_toggle_switch} is mixed with white noise, that is, no correlation at all between the two sources. The IDL function RANDOMU was used for pseudo-random number generation. The intermediate case (shown in the top panel) occurs when enough to make the observers' prediction of the next sample incorrect 25\% of the time. To put this another way, Alice is still certain to receive a blue photon when Bob receives red, on average 3 out of 4 times, and likewise reliably predict each-other's results. They need not know why, whether from intrinsic signal or correlated noise at detection. But noise at the switching telescopes due to local sky conditions, detector noise, etc., if {\it uncorrelated} must work randomly against the accuracy of the colour measurements at either station. So all of that noise at the telescopes are assumed, collectively to be Gaussian; simulated with RANDOMN. The effects of adding one standard deviation of observational noise at each station are illustrated in the bottom two panels of Figure~\ref{figure_simulated_relative_colours}. It is notable that these two cases look very similar, but it will be seen that they are distinguishable.

\begin{figure}
\includegraphics[width=5.5cm]{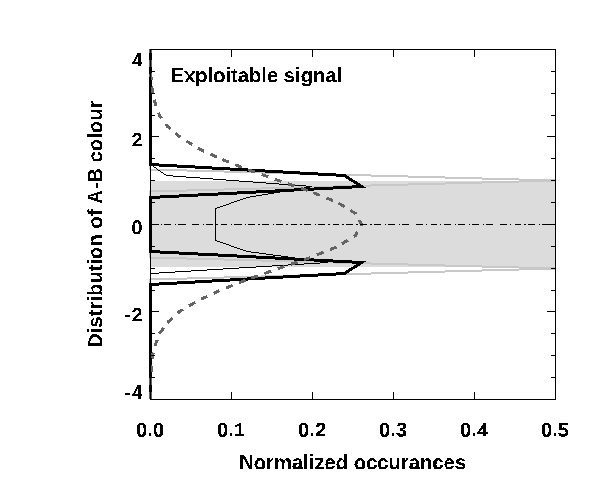}\includegraphics[width=5.5cm]{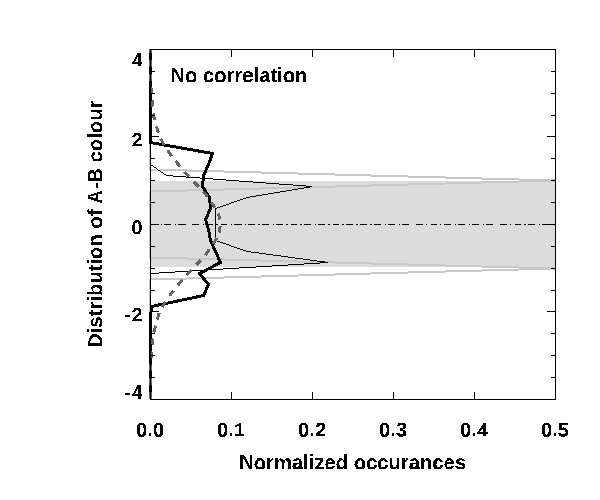}
\includegraphics[width=5.5cm]{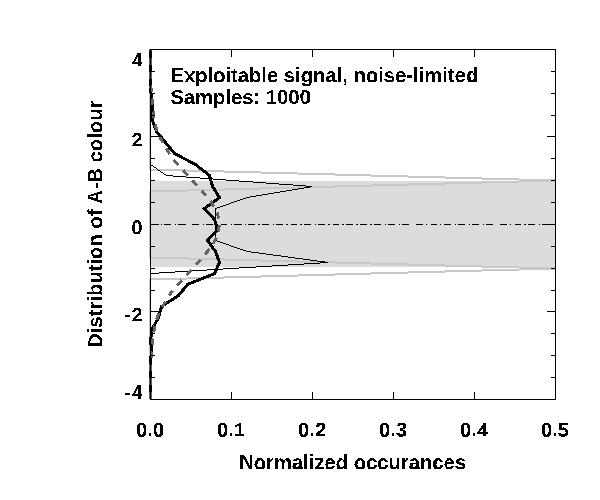}\includegraphics[width=5.5cm]{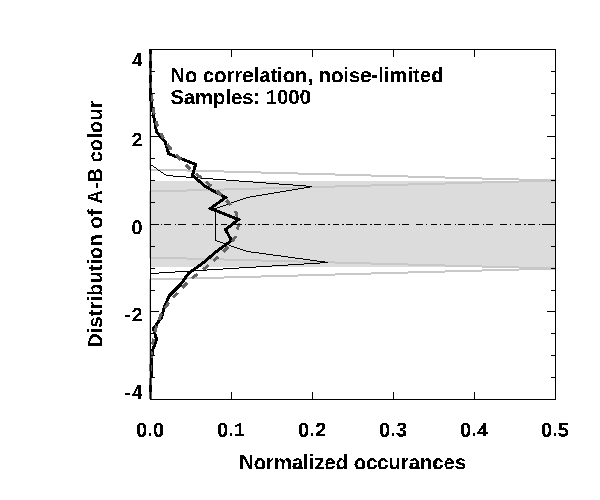}
\includegraphics[width=5.5cm]{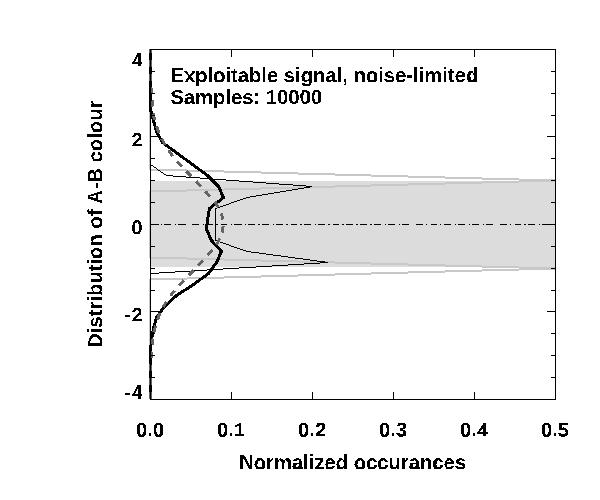}\includegraphics[width=5.5cm]{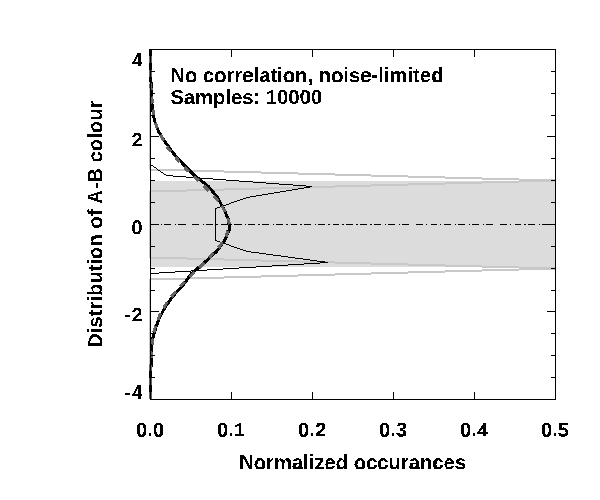}
\caption{Distributions of simulated samples, corresponding to the cases in Figure~\ref{figure_simulated_relative_colours} as thick black curves. The top panels are for 1000 samples without observational noise, either assuming exploitable signal (upper-left panel) or only white-noise, i.e. with no detectable correlation (upper-right panel); the case of a pure sinusoid is indicated as a thin black curve, and the ideal ``toggle switch" in thin light-grey; a span of 2-standard deviations is shaded grey. Below these top panels are those same cases, limited by observational noise; either for 1000 samples (middle panels) or 10000 samples (bottom panels). Notice that a signal just-sufficient to exploit in defeating a QM experiment would remain discernably different from an associated Gaussian-noise distribution, indicated in each panel by a dashed-grey curve; even so, these have the same observational noise added.}\label{figure_simulated_distribution_colours}
\end{figure}

If the exploitable minimally-correlated case does occur it is important to note that this need not be detected in each individual difference; the underlying correlation is relative to the added noise, and statistical. However, the correlation must persist for at least as long as an observation, and have frequency comparable to the sampling period. That is because the distribution of these A-B colours are measurably distinct, which is illustrated in Figure~\ref{figure_simulated_distribution_colours}. The top two panels are those distributions generated from the sequences shown in the top two panels of Figure~\ref{figure_simulated_relative_colours}. The grey shaded band is the same one-standard-deviation range over which the colours must deviate in order to minimally detect correlation. Notice that, in the perfect case of a toggle-switch, every sample at each receiver/switching station would be either blue (one std. dev. up from the mean) or red (one std. dev. down) exactly 50\% of the time, even if not sequentially. In the top-left hand panel it can be seen that this exploitable-signal case (thick black histogram) is no longer perfect, but shares the same attributes of the simple sinusoid (thin black curve). And when degraded to one standard deviation of noise at each observing station, it now offers only a  50:50 prediction of the next outcome (losing its utility to spoil an experiment), but is noticeably not Guassian in distribution. On the contrary, for the cases of no underlying correlation, which are shown in the right hand panels, the distribution of detected colours is clearly Gaussian.

\begin{figure}
\includegraphics[width=9.5cm]{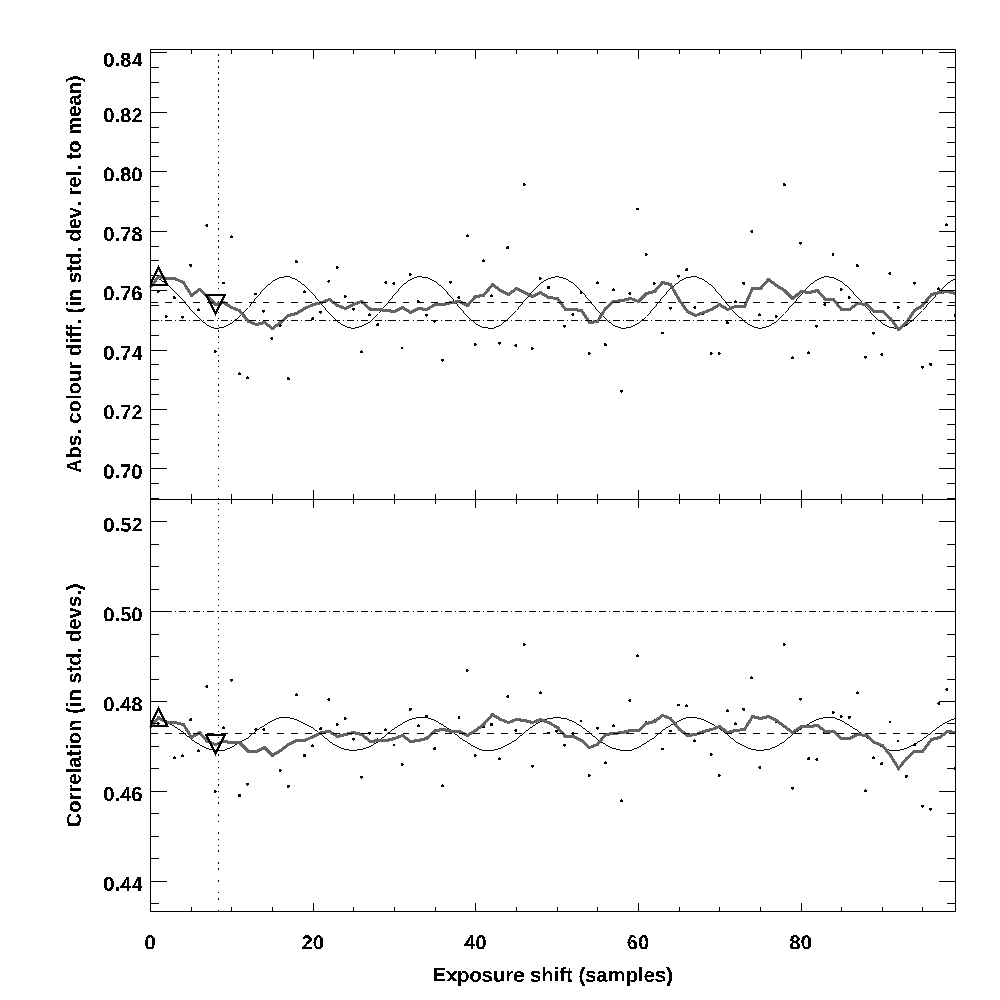}
\caption{Auto-correlation of a 1000-sample simulated sequence, limited by noise; each point is the result of taking the difference after shifting by that number of samples. No correlation is detectable.}\label{figure_simulated_correlations}
\end{figure}

Another way of establishing no exploitable information is the auto-correlation of the signal; shown in Figure~\ref{figure_simulated_correlations}, and obtained by shifting each sample B sequence relative to the A sequence incrementally sample-by-sample and finding the colour difference. If some particular offset in time between the two signals did increase the correlation, this would cause a change relative to the mean colour difference. In this case, there is none, as expected. Note that the absolute difference in colours (in units of std. dev. from the mean) is always greater than 0.75 within some margin of about 0.02 (2\% error); this corresponds to a correlation under 0.50 (again imperfect by the same error margin) equivalent to even odds of prediction of the colour in any sample for either A or B.  To help compare, a vertical dotted line indicates the first 8 samples, with the sinusoidal curve of Subsection~\ref{plausible_signal} overplotted. The upward-pointing and downward-pointing triangles are the first high and last low results within that span. This illustrates that despite the seeming correlation (those do appear to agree with the trend) these samples are random; that signal is not present. It is helpful to look back to Figure~\ref{figure_simulated_relative_colours}; notice that this result corresponds to the lower, right panel: a Gaussian distribution. The case in the lower-left panel is a distinctly bimodal distribution, with just-sufficient noise to render its signal insufficient to defeat a Bell test.

\vspace{3 cm}

\subsection{Incorporating Site Conditions into a Model of Measureable Signal-to-Noise Ratio}\label{site_conditions}

\begin{figure}
\includegraphics[width=9.5cm]{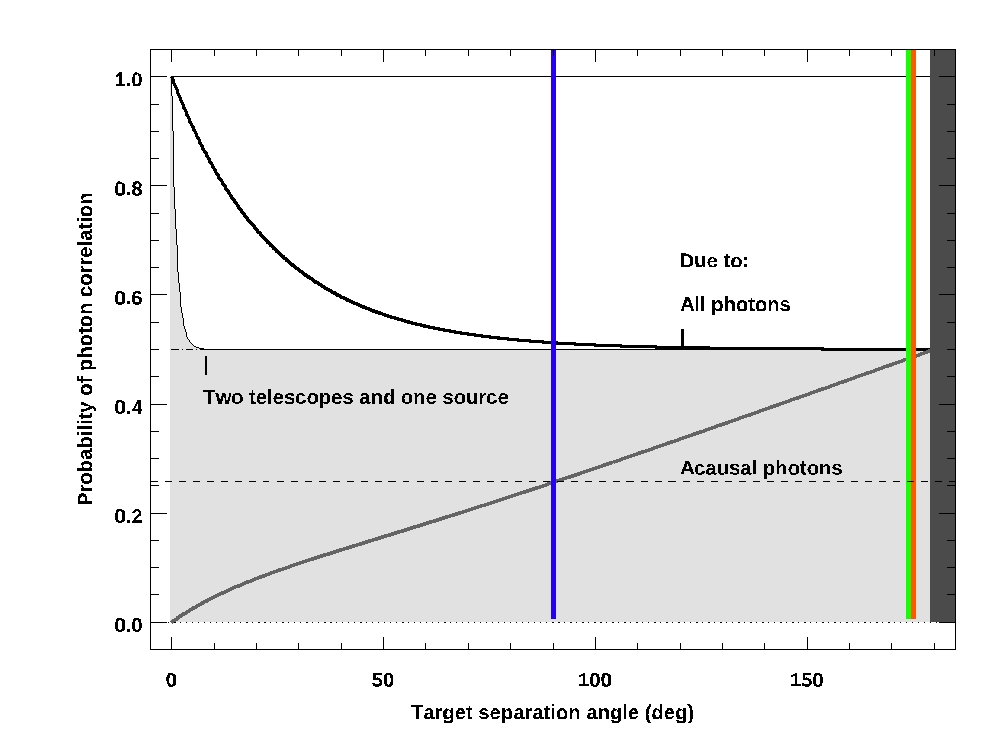}
\caption{Boundary conditions for the probability that detected photons are correlated in equation 2, demanding that the fraction of acausal-pair photons is 100\% at $180^{\circ}$; median situation shaded grey.}\label{figure_fraction_and_probability}
\end{figure}

Whatever the observable A-B colour signal of two pointsources at high angular separation, and acausal, where unseparated it must be the best possible condition: both telescopes pointed at a {\it single} source. In that case, the probability $P$ of those two telescopes sensing perfect correlation among the emitted photons (without noise) is exactly unity: $$S\equiv\omega\times P=\omega, \eqno(1)$$where $\omega$ is the colour (in magnitudes) per sample of the source; the {\it perfectly correlated} difference in photon flux between blue and red filters, that is, the amplitude of the sinusoid in Figure~\ref{figure_schematic_toggle_switch}. The probability of those photons being acausal is exactly zero; they are coming from the same location (and source). This situation is shown in Figure~\ref{figure_fraction_and_probability}; 100\% of photons are acausal when both sources have higher than redshift $z=3.65$ (right of the vertical red line) separated by $180^\circ$. Smaller angles (left of the vertical green line) are sources that do not meet this limit, but are visible from two well-separated sites. A separation of about $90^\circ$ is a practical limit of a single observing site (left of the vertical blue line). In the lower panel are the probabilities of any sample sensing a pair of acausal photons, for exponential decay, $$P = 0.5\dot f \Big{[}1 + {{1}\over{e}}\exp\Big{(}1-e^{\rm fold}{\phi\over{180}}\Big{)}\Big{]}, \eqno(2)$$for $\phi$ separation in degrees, and $e^{\rm fold}$ is the e-folding rate for fraction $f$ of acausal-pair photons. This functional form is convenient and smooth, and as will be shown, falls off faster than observational errors grow. It is correct at the extreme limits of no-separation and $180^{\circ}$; for example, for an e-folding factor of 5, just sufficient to approach 50:50 at $90^\circ$. That is a conservative requirement \cite{Friedman2013}, especially for $z=4$, which is adopted hereafter.

Three different scenarios for viewing two quasars in a QM experiment can now be compared. A single site cannot effectively view quasar-pairs with separations larger than about $90^\circ$ because extinction and seeing worsen quickly (although less than exponentially) for each telescope with zenith airmass $Z(\phi)\approx 1/\cos(\phi)$~\cite{Kasten1965} and so incurs an error that grows with separation up to some limiting observable airmass, typically under 2, as$$\sigma=Z_{\rm limit}\Big{|}1 - \sqrt{2}{{Z(\phi)}\over{Z_{\rm limit}}}\Big{|},\eqno(3)$$for two telescopes at one site. These errors are assumed to be symmetric, without colour bias, thus the absolute value is taken. The combined observational noise increases as$$N=\Big{[}{\sigma(\alpha + {\sigma}^2 \beta {\gamma}^2) + \epsilon \over{\sqrt{n}}}\Big{]} + \zeta, \eqno(4)$$where $\alpha$ is extinction, $\beta$ is brightness of sky beyond darkest possible, $\gamma$ is seeing (all at zenith); $\epsilon$ is photometric error per $n$ samples; and $\zeta$ is the instrumental zeropoint error. Two sites separated by $90^\circ$ allow larger viewable separations, because zenith angle now has form $Z(\phi/2)$. And third: two telescopes at antipodal sites on the Earth, which has the form $Z(180^\circ-\phi)$, and might include Hawai'i and southern Africa, for example. Such mid-latitude locations have 8-m-class telescopes, but do not allow simultaneous viewing of the {\it same} location on the sky, so must incur at least a penalty of $\sqrt{2}\zeta$ in zeropoint calibration.

\begin{table}[H]
\caption{Values of Parameters Used in Model}
\begin{tabularx}{\textwidth}{ccccccc}
\toprule
Samples &Signal   &Extinction &Sky brightness par.             &Seeing   &Photom. err. &Zeropoint err.\\   
$n$     &$\omega$ &$\alpha$   &$\beta$                         &$\gamma$ &$\epsilon$   &$\zeta$       \\
        &(mag)    &(mag)      &(${\rm mag}~{\rm arcsec}^{-2}$) &(arcsec) &(mag)        &(mag)         \\
\midrule
3000    &0.02     &0.25       &0.50                            &0.80     &0.10         &0.01          \\
\bottomrule
\end{tabularx}
\label{table_model_parameters}
\end{table}

\begin{figure}
\includegraphics[width=9.5cm]{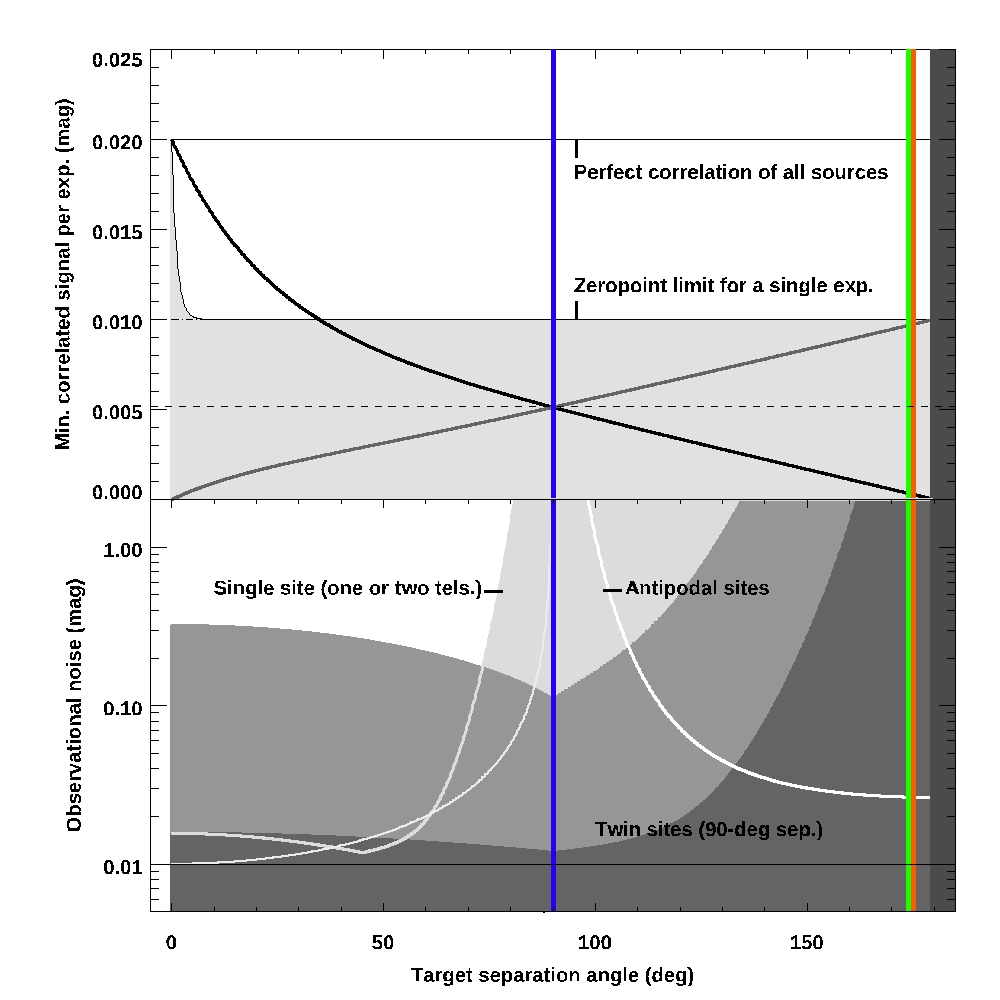}
\caption{Minimally-detectable correlated signal per sample (top panel) for identical instruments; and total observational noise (bottom panel), including zeropoint, photometric error per sample, seeing and sky background for two telescopes at one site, and two sites separated by 90 degrees on the Earth (single exposure: light grey, 3000 samples: dark grey) or two sites at antipodes (thick white curve).}\label{figure_noise_and_signal}
\end{figure}

Imagine the smallest-detectable correlation between two sources, that is when $\omega=2\zeta$, and the A-B colour difference per sample is at twice the zeropoint limit of each instrument; ideally identical. Signal-per-exposure as a function $\phi$ (with parameters as in Table~\ref{table_model_parameters}) are plotted in Figure~\ref{figure_noise_and_signal}: signal per exposure (top panel) is set by requiring that the on-axis (unseparated) sources cannot differ by more than this smallest measureable signal; here, in magnitude units. The total observational noise, including zeropoint, background and seeing for two sites separated by 90 degrees on the Earth are also shown (bottom panel); other geometries indicated as shaded regions, i.e. two telescopes at one site, or two sites at antipodes (white curve) incuring twice the zeropoint noise, as they cannot be directly cross-calibrated. No observational scenario could do better than about $0.01~{\rm mag}$ noise per sample due to the zeropoint of real systems (roughly 1\% photometry) even at a single site. A tradeoff (not shown) is for sites at separation larger than 90 degrees, compromising between those two previous scenarios. This last scenario is that of Gemini Observatory, with twin telescopes situated at a separation of about $95.5^\circ$ in great-circle distance.

\section{Observations}\label{data}

\begin{figure}
\includegraphics[width=13.5cm]{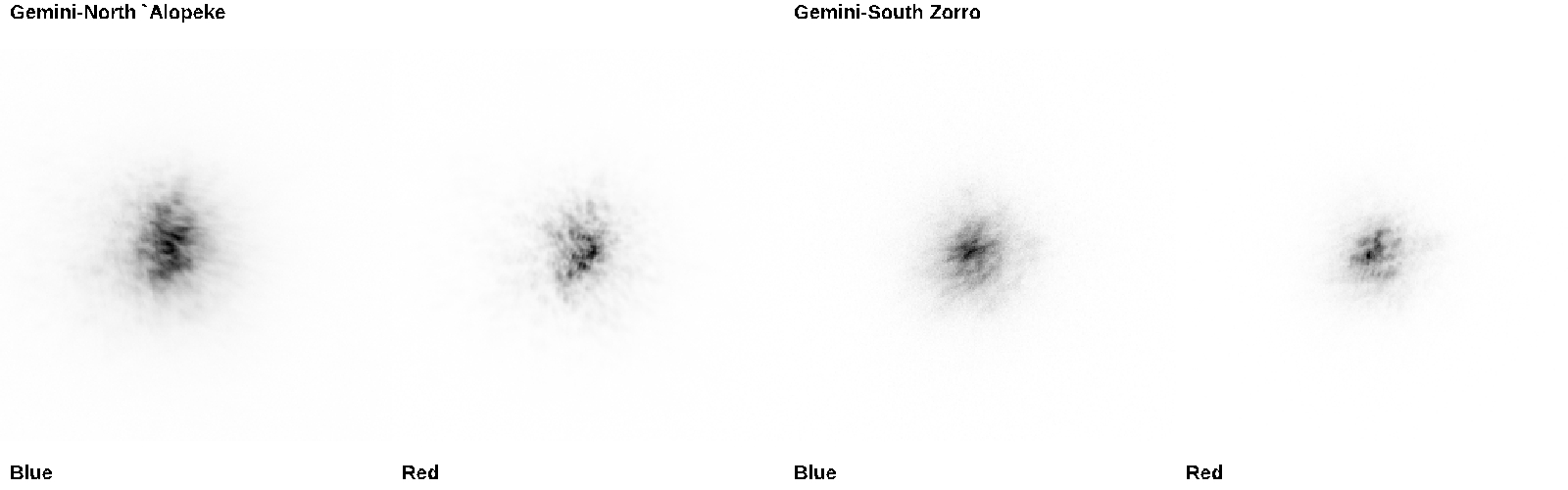}
\caption{Sample images from the Gemini `Alopoke/Zorro dataset; in blue and red filters. These are simultaneous $0.06~{\rm s}$ integrations of two bright stars, one from the North and one from the South.}\label{figure_Alopeke_Zorro_frames}
\end{figure}

Gemini provides a suitable dataset for demonstration of the search, by operating identical high-framerate (up to $67~{\rm Hz}$) low-noise 1024 by 1024 Electron-Multiplying Charge-Coupled Device (EMCCD) optical imagers `Alopeke and Zorro: dual-channel photometers; splitting flux into a blue (centred at $562~{\rm nm}$, $54~{\rm nm}$ wide) and red ($832~{\rm nm}$, $40~{\rm nm}$) passbands \cite{Scott2021}. Installed in 2018, these are normally utilized (independently) in speckle interferometry, although the seeing-limited raw images are each recorded uncorrected. A typical set of 4 images is shown in Figure~\ref{figure_Alopeke_Zorro_frames}. Observations with either instrument are normally a $60~{\rm s}$ long sequence of 1000 exposures; the fast readout electronics allows acheiving nearly $0.06~{\rm s}$ integrations, each timestamped to a within microseconds. The absolute accuracy of the sequences are better than $1~{\rm s}$. The public archive was searched for serendipitous, simultaneous observations when both `Alopeke and Zorro obtained sequences starting within 5 seconds of each other. In all, 4 cases of three sets of observations of two bright stars (a total of 12 observations, listed by start time in Table~\ref{table_observations}) were found that met these timing restrictions, with assurance that a loss of no more than 10\% of the combined dataset. Equivalently, each has more than a 90\%-overlapping period of simultaneous observations.

\begin{table}[H]
\caption{Stars Imaged Simultaneously for 3 Minutes Each with Gemini `Alopeke and Zorro}
\begin{tabularx}{\textwidth}{llrrccrl}
\toprule
           &                 &R.A.    &Decl.    &V     &Sep.  &             &Time   \\
Instrument &Target name      &(deg)   &(deg)    &(mag) &(deg) &Start date   &(UT)   \\
\midrule
`Alopeke   &HR8149           &315.314 &+11.2033 &5.97  &-     &9 Oct. 2019  &7:39:19\\
Zorro      &VZ Col           & 75.503 &-42.6972 &9.71  &119.6 &             &7:39:15\\
           &                 &        &         &      &      &             &       \\
`Alopeke   &CG Peg           &315.687 &+24.7731 &11.2  &-     &20 Oct. 2021 &5:40:45\\
Zorro      &HR451            & 15.577 &-15.6761 &5.61  &71.0  &             &5:40:46\\
           &                 &        &         &      &      &             &       \\
`Alopeke   &G214-012         &330.161 &+41.0363 &13.1  &-     &21 Oct. 2021 &7:52:52\\
Zorro      &HR1870           & 75.526 &-45.9250 &5.86  &127.6 &             &7:52:57\\
           &                 &        &         &      &      &             &       \\
`Alopeke   &TYC 4005-999-1   &345.946 &+55.9772 &11.7  &-     &21 Oct. 2021 &8:35:54\\
Zorro      &UCAC4 505-016395 & 75.853 &+10.8006 &12.5  &81.0  &             &8:35:50\\
\bottomrule
\end{tabularx}
\label{table_observations}
\end{table}

\section{Analysis and Results}\label{results}

Aperture photometry was carried out on all images; a 5-arcsec synthetic aperture was applied, with a 1-arcsec-wide annulus surrounding it to obtain a sky estimate. A custom IDL code was written to perform this, which also generated the statistics for the combined data over the full 60 seconds of each sequence. The difference in flux for blue-red is the colour; this was converted to an AB magnitude via the published bandpasses and throughputs for the instruments, close to 90\% for blue and 70\% in red. 

\begin{figure}
\includegraphics[width=14cm]{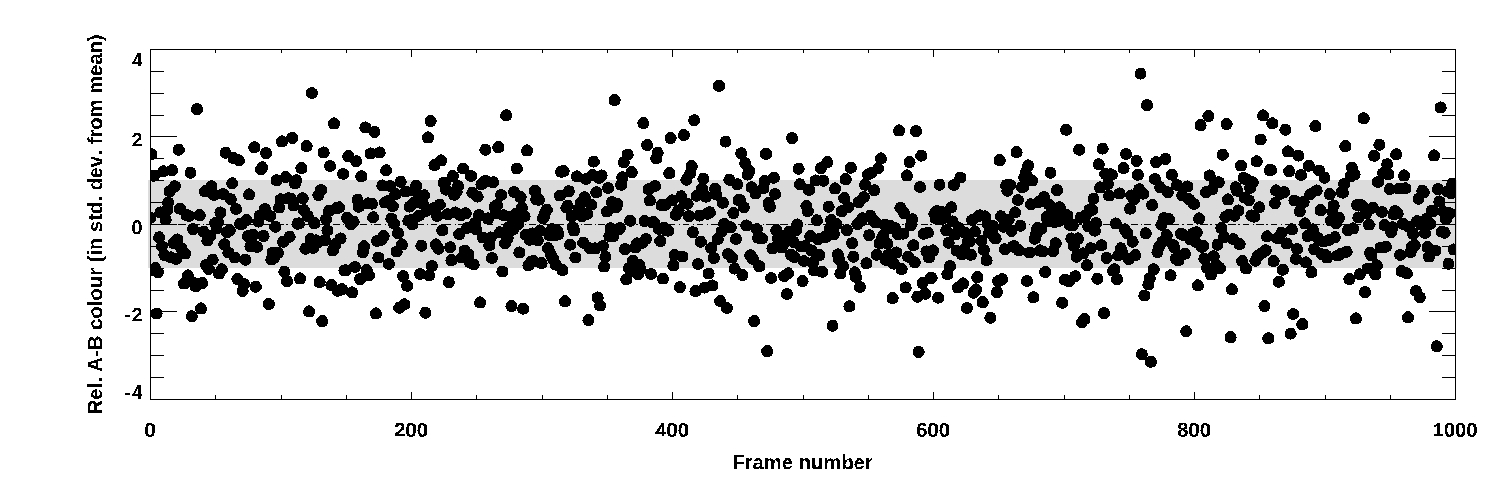}
\caption{Sample A-B colour photometry for one 1000-sample sequence of `Alopeke/Zorro data. These are the first sequence from the data of 9 October 2019.}\label{figure_Alopeke_Zorro_relative_colours}
\end{figure}

\begin{figure}
\includegraphics[width=5.5cm]{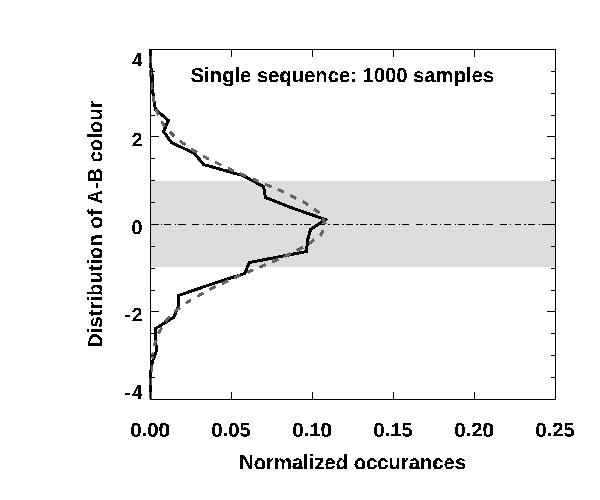}\includegraphics[width=5.5cm]{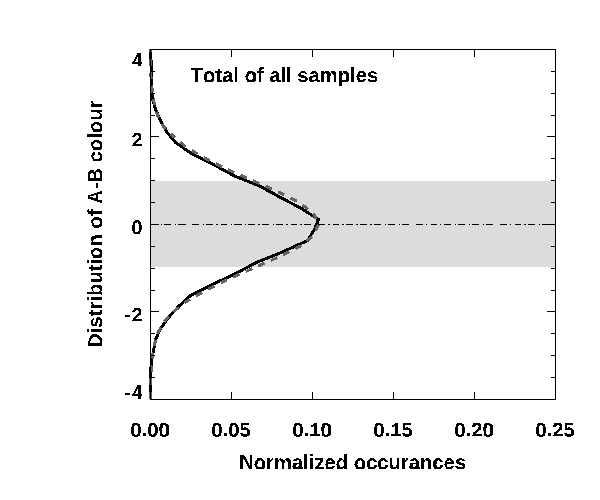}
\caption{Colour distributions for `Alopeke/Zorro data; for the 1000-frame sequence in Figure~\ref{figure_Alopeke_Zorro_relative_colours} (left panel) and for all 12 available archival observations (right panel) for a total of 12000 samples.}\label{figure_Alopeke_Zorro_distributions}
\end{figure}

This photometry was then converted to a difference relative to the standard deviation of colour for the whole sequence; hence, reporting it in std. dev. from the mean, as defined in Section~\ref{plausible_signal}. Results for one sequence are shown in Figure~\ref{figure_Alopeke_Zorro_relative_colours}, that is the A-B colour of the two stars in each frame. The convention chosen was that object labelled A was that observed with `Alopeke, and B for Zorro. A linear fit for colour over the sequence was also subtracted to ensure no trend due to airmass change, although this was found to be a negligible effect, well under 1\%.  The grey band in Figure~\ref{figure_Alopeke_Zorro_relative_colours} delineates 1 std. dev. above and below the slope-corrected mean. The resulting distributions are shown in Figure~\ref{figure_Alopeke_Zorro_distributions}; both one individual sequence of 60 seconds (left: from the same sequence as shown in Figure~\ref{figure_Alopeke_Zorro_relative_colours}), and the combination of all 12 available observations (right), both of these are indistinguishable from Gaussian (dashed curve) normalized to the same peak.

\begin{figure}
\includegraphics[width=9.5cm]{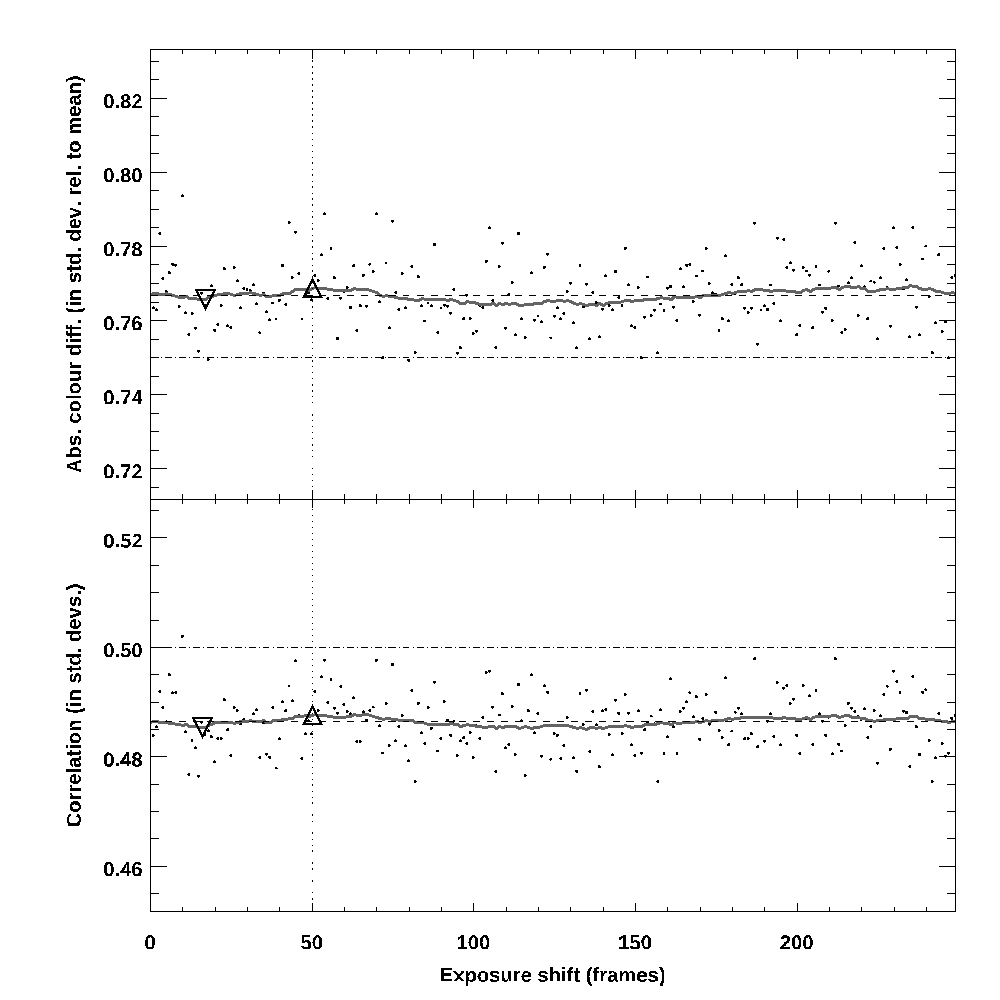}
\caption{Auto-correlation of `Alopeke/Zorro data in Figure~\ref{figure_Alopeke_Zorro_relative_colours}. No detectable correlation is seen, allowing shifts of up to $5~{\rm s}$ for different start times (Zorro precedes `Alopeke), and for clock accuracy.}\label{figure_Alopeke_Zorro_correlations}
\end{figure}

An auto-correlation analysis identical to that described in \ref{autocorrelation} was also carried out, which confirms the random nature of the colours observed, shown in Figure~\ref{figure_Alopeke_Zorro_correlations}. Shifts up to 5 seconds were allowed; the vertical dotted line indicates time period by which the two local instrument clocks could be out of sync; there is no evidence of improved correlation within that window. It perhaps comes as no surprise that the simultaneous colours of these stars are random, but it is still a useful excercise. That is because, although these observations were undertaken at 0.06-second cadence, `Alopeke/Zorro are capable of 0.01-second exposures - shorter than the telescope-to-telescope light-travel time - which it will be shown in the next section can allow a test for two quasars, both much fainter, but still providing sufficient photons to image at that highest framerate.

\subsection{Finding Quasar Pairs to Complete a True Acausal-Photons Correlations Test}\label{quasar_pairs}

A software tool was developed to predict where pairs of quasars with the potentially-observable, and exploitable A-B colour signal of Section~\ref{plausible_signal} might be found. It incorporates expected noise limitations imposed at different sites and instrument properties (telescope aperture, framerate and throughputs) as in Subsection~\ref{site_conditions}, and then reports a SNR for those selected quasar pairs, based on catalogued brightnesses. The code can be set for Gemini (8.1-m apertures, geographic coordinates) and calculates the visibility of sources for the upcoming year and outputs a target list indicating the best night during that time. It can also restrict allowed sky-offsets from a single object (for example, a calibration star) to be simultaneously visible from both sites on that night. The best time in the night is when both targets are at the lowest combined airmass for both sites: where both sources reach their highest ascension in both North and South skies. Outputs are target names for each of the two selected telescope sites, appending A and B with the convention that target A is rising and B setting on the best night. This program is called PDQ (Predict Different Quasars); written in IDL, and is freely available via GitHub. The target list remains unvetted in this version of the code: neither the brightness of the quasars, nor their redshifts are verified. A sample is listed in Table~\ref{table_quasar_pairs}. Manual checking with slightly relaxed settings does find some suitable pairs bright enough to image with `Alopeke/Zorro.

\begin{table}[H]
\caption{Potential Acausal Quasar Pairs Visible for Over Two Weeks with Gemini `Alopeke and Zorro}
\begin{tabularx}{\textwidth}{lcccccc}
\toprule
                             &R.A.    &Decl.    &V     &     &Separation &            \\
Target name                  &(deg)   &(deg)    &(mag) &$z$  &(deg)      &Best date   \\
\midrule
SDSS J104837.40-002813.6 (A) &150.810 &+0.47044 &18.3  &4.03 &-          &30 May 2025 \\
SDSS J225928.76-015623.0 (B) &330.991 &-1.93972 &18.3  &4.3  &178.5      &            \\
                             &        &         &      &     &           &            \\
SDSS J222123.72-004345.1 (B) &330.357 &+0.72919 &15.3  &4.3  &-          &27 Nov. 2025\\
SDSS J104837.40-002813.6 (A) &150.810 &+0.47044 &18.3  &4.03 &178.7      &            \\
                             &        &         &      &     &           &            \\
SDSS J233122.06+074811.7 (B) &345.523 &+7.80325 &18.4  &4.9  &-          &13 Dec. 2025\\
SDSS J113723.38-060217.0 (A) &165.623 &-6.03806 &18.4  &4.1  &178.2      &            \\
                             &        &         &      &     &           &            \\
SDSS J233122.06+074811.7 (B) &345.523 &+7.80325 &18.5  &4.9  &-          &14 Dec. 2025\\
BR 1144-0723 (A)             &165.777 &-7.66806 &18.5  &4.16 &179.7      &            \\
\bottomrule
\end{tabularx}
\label{table_quasar_pairs}
\end{table}

\begin{figure}
\includegraphics[width=9.5cm]{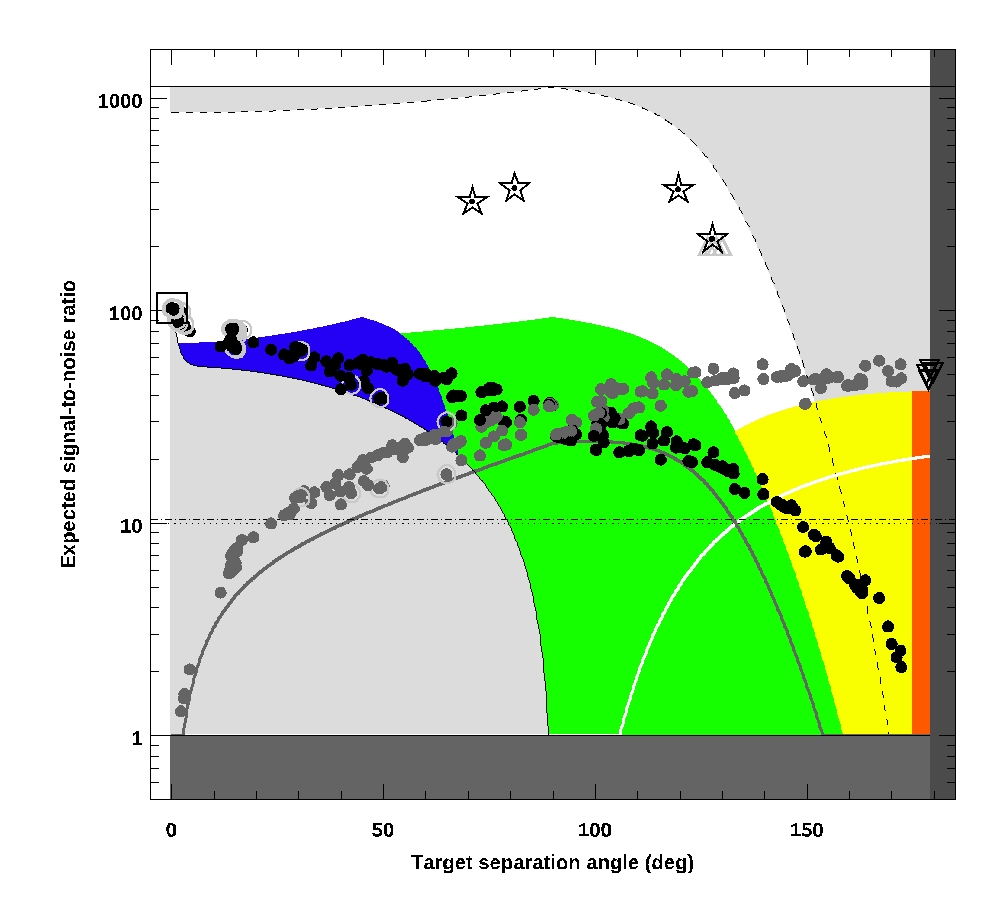}
\caption{Predicted A-B colour SNR for acausal photons (filled grey circles) from quasar pairs with Gemini  `Alopeke and Zorro, as found with the PDQ software, plotted for instances when objects are visible for at least two weeks; same, for the causal fraction of photons (filled black circles); down-pointing triangles are for $z\geq 4$, $V=18.5~{\rm mag}$ acausal quasar pairs; for comparison, up-pointing triangles are a transitional case at peak SNR photons (but not fully acausal) near $130^{\circ}$ separation, for 10000 samples of $V=15.5~{\rm mag}$, $z=1.48$ quasars; open circles flag those cases which could be observed by either site; an open square indicates both telescopes observing one source (causal, and the same brightness as the acausal pairs); open stars are the stellar sources already observed by Gemini for 3000 samples each. The outer grey-shaded boundary is the model result for fair-sampling (indistishible between causal and acausal photons; equation 2 with $f=0.5$) and $90^{\circ}$-separated sites and $V=12~{\rm mag}$ sources; similarly for $V=18.5~{\rm mag}$ sources (green), a single site (blue), and two antipodal sites (yellow) via the model of Section~\ref{methods}. Notice that on optimal nights, Gemini allows conditions for viewing fully acausal sources (red), with SNR similar to antipodal sites.}\label{figure_ratio}
\end{figure}

\begin{figure}
\includegraphics[width=11cm]{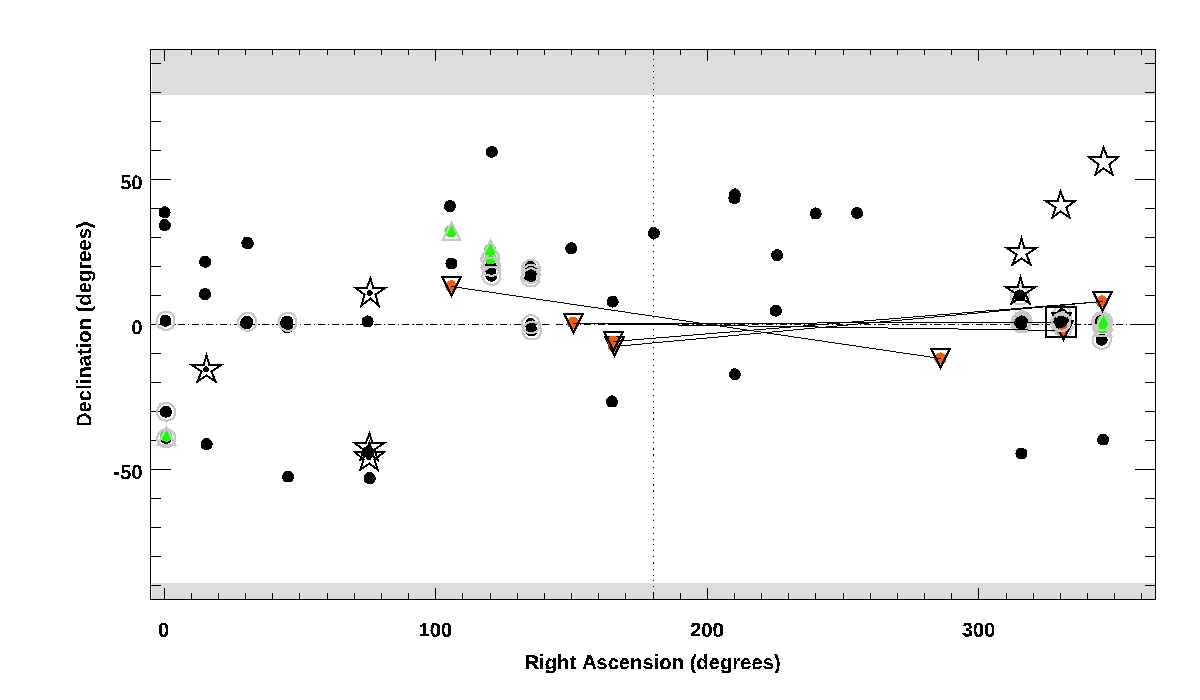}
\caption{Source pairs on the sky; symbols are as in Figure~\ref{figure_ratio}; here, the acausal quasar pairs are connected by a thin black line segmment; the South (B) star of pair is indicated by a central black dot.}\label{figure_sky}
\end{figure}

\begin{figure}
\includegraphics[width=7cm]{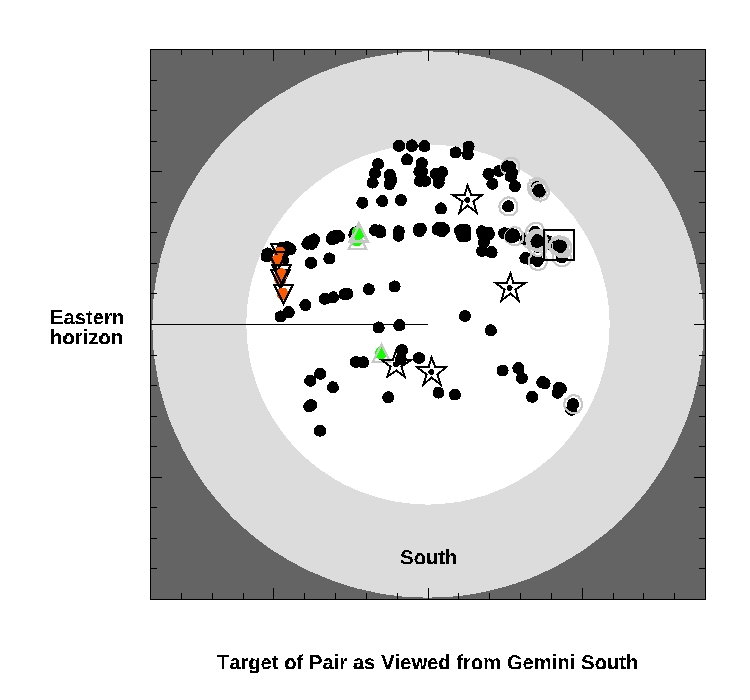}
\includegraphics[width=7cm]{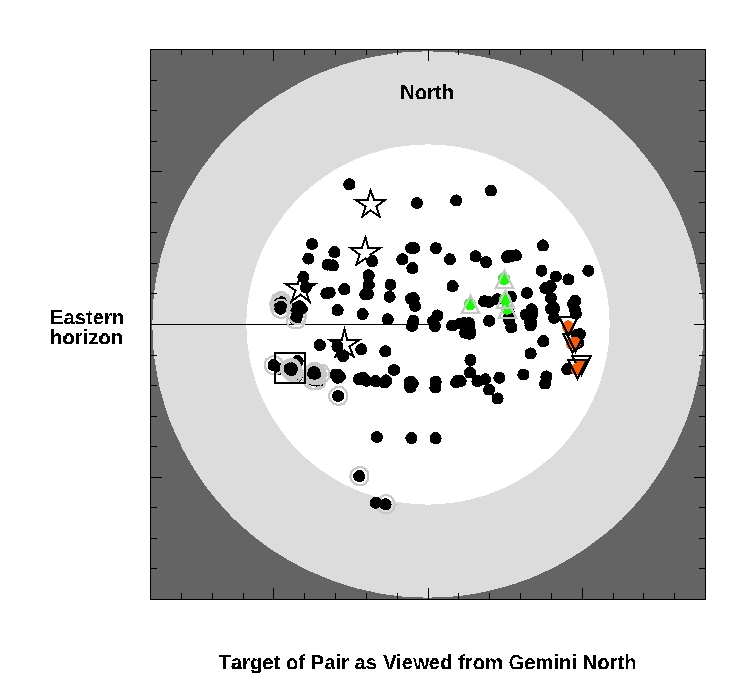}
\caption{Same as Figure~\ref{figure_sky}, in polar projection from zenith, as each is best viewed from either Gemini North or South, at lowest combined airmasses: a preferential East-West alignment evident.} \label{figure_polar}
\end{figure}

Using the results from Section~\ref{results} on bright stars for comparison, the tool's output is plotted in Figure~\ref{figure_ratio}: A-B observed-colour SNR for 3000 samples at 50 Hz. Alopeke/Zorro results are indicated by black-outlined stars; all others are for 18.5 R-mag and brighter quasars selected from the MILLIQUAS sample \cite{Flesch2023}; either of $z=1.48$ (the mean sample reshift) and separated by $130^\circ$; or requiring that each has a (possibly photometric) redshift of $z\geq4$; those of separation $180^\circ$ are indicated by open down-pointing triangles. At all smaller separation angles these are indicated by the minimal fraction of acausal photons per sample (dark grey filled circles) and causal (black). Where these would be visible to both telescopes, these are outlined by a grey circle; a black square indicates when it is, in fact, the same source: zero separation. Coloured and shaded regions indicate expected model-SNR limits for fair-sampling at other telescope-site choices (equation 2 with $f=0.5$): light grey is a limiting case if a single telescope could have a field of view over the whole sky; blue is for a single site with two telescopes; green indicates two sites of exactly $90^\circ$ separation on the Earth; and yellow for antipodal sites; an upper limit (for $90^\circ$ separation) from the model at 50 Hz, but with flux matching that of the observed stars (dashed curve).

Two further outputs help with visualizing the orientation of the potential targets on the sky, and possible scheduling of observations: Figure~\ref{figure_sky} is a plot of right ascension and declination of the targets displayed in Figure~\ref{figure_ratio}; as those visible in the coming year. Note the East-West alignment of acausal quasar-pairs that are separated by $180^\circ$ (down-pointing tringles, with red filled circles). For comparison, the already-observed stars are shown; a black central dot indicates where viewed from Gemini-South. In Figure~\ref{figure_polar}, polar projections of the visibility of the targets are shown for Gemini-South (left) and Gemini-North (right); again, it is clear that suitable acasual quasar pairs are visible towards the low eastern horizon (Gemini-South) and western sky (Gemini-North) simultaneously.

\section{Summary and Conclusions}\label{conclusions}

True random-number generation can be demonstrated via high-framerate photometric observations of quasar pairs that are separated by angles on the sky that make them inaccessible to any single ground-based observatory site. When $180^\circ$ apart, and each at $z\geq3.65$ their emitted photons are acausal: no signaling between the sources could have spoiled the independence of a Bell-test setting procedure. So far, no such observations have been undertaken, and thus that is not proved. An analysis has been carried out that sets boundaries on how observably correlated those two sources might be; and the statistical signature of detecting that correlation relative to local noise sources. Simulated data were generated, and then an analytic model used to compare among telescopes in three potential observing scenarios, and for any two observatory sites. It is found that two 8-m-class telescopes at good sites at over $90^\circ$ separation on Earth are able, for reasonable observational limits and choices of optical bandpasses, and at high but practical camera framerates of $50~{\rm Hz}$ to $100~{\rm Hz}$ to achieve suitable SNR to overcome seeing and sky-noise conditions, and so rule out correlated noise ruining observations of suitable quasar pairs in QM tests. Details of experimental procedures are left for other work.

To demonstrate the observational method, and the utility of Gemini for the task, archival observations of 4 sets of bright star pairs obtained simultaneously (and serendipitously) in two bandpasses with `Alopeke/Zorro were presented. These do not meet the framerate required, obtained at only $17~{\rm Hz}$, but the necessary sampling speed is obtainable with this instrumental setup at its highest setting, $67~{\rm Hz}$. That also provides a calibration of the SNR achievable; the model described is able to extrapolate to fainter quasars observed at that framerate: typical sky conditions at the Gemini sites suggest acausal 18.5-mag $z\geq4$ quasars can achieve SNR of 50 in 3000 samples, ruling out correlation sufficient to spoil a QM test lasting minutes. And finally, a tool for finding suitable quasars to carry out such an observation and confirm no detectable correlation at the required framerate was presented; with example results of running this PDQ code set for observations with Gemini `Alopeke/Zorro. It is found that several potential quasar pairs could be visible in the coming semesters, and so actual observations with Gemini may be considered. In the meantime, that code is freely provided via GitHub for the community, for example, in planning further QM tests. 


\vspace{6pt}

\funding{This research received no external funding.} 

\dataavailability{All data discussed herein are freely available via public archives using the links indicated in the text, and their references. Those data are also all downloadable from \url{https://github.com/ericsteinbring/PDQ} in formatted data tables, together with the analysis code, written in IDL. In its default settings, that code generates the figures shown here.} 




\acknowledgments{I gratefully acknowledge helpful comments from the anonymous referees, which improved the manuscript. This research used the facilities of the Canadian Astronomy Data Centre operated by the National Research Council of Canada with the support of the Canadian Space Agency. Archival observations employed the High-Resolution Imaging instruments ‘Alopeke and Zorro, funded by the NASA Exoplanet Exploration Program and built at the NASA Ames Research Center by Steve B. Howell, Nic Scott, Elliott P. Horch, and Emmett Quigley. ‘Alopeke and Zorro were mounted on the Gemini North and South telescopes of the international Gemini Observatory, a program of NSF NOIRLab, which is managed by the Association of Universities for Research in Astronomy (AURA) under a cooperative agreement with the U.S. National Science Foundation on behalf of the Gemini partnership: the U.S. National Science Foundation (United States), National Research Council (Canada), Agencia Nacional de Investigación y Desarrollo (Chile), Ministerio de Ciencia, Tecnología e Innovación (Argentina), Ministério da Ciência, Tecnologia, Inovações e Comunicações (Brazil), and Korea Astronomy and Space Science Institute (Republic of Korea).}

\conflictsofinterest{The author declares no conflict of interest.} 



\abbreviations{Abbreviations}{
The following abbreviations are used in this manuscript:\\

\noindent 
\begin{tabular}{@{}ll}
AGN & Active Galactic Nucleus\\
IDL & Interactive Data Language\\
NIR & Near Infrared\\
QM & Quantum Mechanics\\
SNR & Signal-to-Noise Ratio
\end{tabular}
}




\begin{adjustwidth}{-\extralength}{0cm}

\reftitle{References}

\PublishersNote{}
\end{adjustwidth}
\end{document}